\documentclass[prd, twocolumn,nofootinbib,10pt]{revtex4-1}
\date{\today}

\usepackage{epsfig}
\usepackage{amsmath}
\usepackage{amsfonts}
\usepackage{graphicx}
\usepackage{dsfont}
\usepackage[german, english]{babel}
\usepackage{amssymb}
\usepackage{ifthen}
\usepackage{ulem}
\usepackage{color}
\usepackage{float}

\newcommand{\secref}[1]{Section \ref{#1}}
\newcommand{\figref}[1]{Figure \ref{#1}}
\renewcommand{\eqref}[1]{Eq. (\ref{#1})}

\usepackage{caption}
\newcommand{\insertplot}[5]{\begin{figure}
		\hfill\hbox to 0.05in{\vbox to #5in{\vfill
				\inputplot{#1}{#4}{#5}}\hfill}
		\hfill\vspace{-.1in}
		\caption{#2}\label{#3}
\end{figure}}
\newcommand{\inputplot}[3]{
	\special{ps: plotfile #1}
}
\newcounter{fig}

\newcommand{\bea}{\begin{eqnarray}}
\newcommand{\eea}{\end{eqnarray}}
\newcommand{\be}{\begin{equation}}
\newcommand{\ee}{\end{equation}}

\newcommand{\re}[1]{(\ref{#1})}

\newcommand{\e}{\mathrm{e}}

\newcommand{\ii}{\mathrm{i}}

\newcommand{\tr}{\mbox{tr}}

\RequirePackage[colorlinks=true,allcolors=blue,
unicode=true,hypertexnames=false]{hyperref}
\hypersetup{pdfstartview={XYZ null null 1.25}}

\definecolor{ao(english)}{rgb}{0.0, 0.5, 0.0}

\newcommand{\cc}{\text{c.c.}}
\newcommand{\FF}{\text{FF}}
\newcommand{\LO}{\text{LO}}
\renewcommand{\U}{\text{U}}

\renewcommand{\max}{\text{max}}
\renewcommand{\min}{\text{min}}

\usepackage{bm}
\renewcommand{\vec}[1]{\bm{#1}}

\usepackage{soul}

\graphicspath{{images/}}

\usepackage{float} 
\usepackage{caption} 
\usepackage{subcaption} 

\usepackage{xcolor}
\usepackage{soul}

\usepackage{bm}
\renewcommand{\vec}[1]{\bm{#1}}
\newcommand{\integral}[2]{\int _{#1}^{#2}} 
\newcommand{\diff}[1]{\text{d}#1} 
\newcommand{\diffD}[2]{\text{d}^{#2}#1} 
\newcommand{\grad}[1]{\nabla #1}
\newcommand{\laplace}[1]{\Delta #1}

\newcommand{\F}{\mathcal{F}}

\renewcommand{\Re}{\text{Re}}

\newcommand{\Fermi}{\text{F}}
\newcommand{\bulk}{\text{bulk}}

\begin{document}
	
\title{Pair-density-wave superconductivity of faces, edges and vertices in systems with imbalanced fermions}

 \author{Albert Samoilenka}
 \affiliation{Department of Physics, Royal Institute of Technology, SE-106 91 Stockholm, Sweden}
 
 \author{Mats Barkman}
 \affiliation{Department of Physics, Royal Institute of Technology, SE-106 91 Stockholm, Sweden}

 \author{Andrea Benfenati} 
 \affiliation{Department of Physics, Royal Institute of Technology, SE-106 91 Stockholm, Sweden}

 \author{Egor Babaev}
 \affiliation{Department of Physics, Royal Institute of Technology, SE-106 91 Stockholm, Sweden}

\begin{abstract}
We describe boundary effects in superconducting systems with Fulde-Ferrell-Larkin-Ovchinnikov (FFLO) superconducting instability, using Bogoliubov-de-Gennes and Ginzburg-Landau (GL) formalisms. First, we show that in dimensions larger than one the standard GL functional formalism for FFLO superconductors is unbounded from below. This is demonstrated by finding solutions with zero Laplacian terms near boundaries. We generalize the GL formalism for these systems by retaining higher order terms. Next, we demonstrate that a cuboid sample of a superconductor with imbalanced fermions at a mean-field level has a sequence of the phase transitions. At low temperatures it forms Larkin-Ovchinnikov state in the bulk but has a different modulation pattern close to the boundaries. When temperature is increased the first phase transition occurs when the bulk of the material becomes normal while the faces remain superconducting. The second transition occurs at higher temperature where the system retains superconductivity on the edges. The third transition is associated with the loss of edge superconductivity while retaining superconducting gap in the vertices. We obtain the same sequence of phase transition by numerically solving the Bogoliubov-de Gennes model.
\end{abstract}

\maketitle

\section{Introduction}

 Fulde and Ferrell \cite{FuldeFerrell1964} and Larkin and Ovchinnikov \cite{LarkinOvchinnikov1964} (FFLO)
 considered a superconducting  state where, Cooper pair forms out of two electrons with 
 different magnitude of momenta.
The original considerations expected such a situation to 
 arise in the presence of a strong magnetic field and thus Zeeman splitting of the Fermi surfaces 
for spin up and spin down electrons.
Later it was shown that in other physical systems the fermionic imbalance occurs without  any applied
magnetic field. Indeed the FFLO state was discussed 
 in cold-atom gases where one can create different imbalances
 of fermions \cite{Zwierlein_2006,Radzihovsky2011,Kinnunen2018,samoilenka2019synthetic}.
 Similarly the difference in Fermi surfaces naturally occurs in dense quark matter.
 The resulting superconducting states of quarks are called 
 color superconductivity which is suggested to realize FFLO state in the  cores of neutron stars \cite{alford2001crystalline}.
 Even in electronic superconductors,   finite momentum pairing may arise due to reasons other than
 application of an external magnetic field \cite{barkman2019antichiral}. 
 This state  has for a long time been of great interest and was searched for in a number of superconducting materials  \cite{Bianchi2003,Ironbased,Uji2018,Mayaffre2014,coniglio2011superconducting,norman1993existence,cho2011anisotropic,matsuda2007fulde,lortz2007calorimetric,singleton2000observation,martin2005evidence}.

In the recent work \cite{barkman2018surface}, using microscopically derived Ginzburg-Landau (GL) model, it was shown that 
systems that support FFLO superconductivity in the bulk do not undergo a direct superconductor-normal metal phase transition.
Instead these systems have a different intermediate phase at elevated temperatures where superconductivity occurs only on the boundary but the bulk of the systems is a normal metal.

In this work we answer two questions. (i) We demonstrate this effect in microscopic models for imbalanced fermions   without relying on a Ginzburg-Landau expansion. We show that in dimensions larger than one, the superconductor undergoes a sequence of phase transition where superconductivity survives in sub-domains of sequentially lower dimension. That is, by increasing the temperature and fermionic population imbalance in a square system, at the mean-field
level, the system will undergo the following sequence of the phase transitions:  $ \text{superconducting bulk}  \to \text{superconducting edges} \to \text{superconducting corners}  \to \text{normal state}$.
(ii) For a Ginzburg-Landau approach, we explain that, as alluded in \cite{barkman2018surface}, in dimensions larger than one, it is necessary to modify the Ginzburg-Landau expansion compared to what is done in the standard calculations \cite{Buzdin1997}  to describe systems with boundaries. 
By retaining higher order derivative terms, we demonstrate the same sequence of phase transition obtained in the microscopic model. That is, a three dimensional cuboid superconductor with imbalance fermions undergoes the sequence of phase transitions: $ \text{superconducting bulk}  \to \text{superconducting faces} \to \text{superconducting edges} \to \text{superconducting vertices}  \to \text{normal state}$, as temperature and fermionic population imbalance is increased.
We also demonstrate an alternative, simpler  Ginzburg-Landau expansion in the presence of boundaries, which also exhibits the boundary pair-density-wave  state, but does not capture the same sequence of phase transitions.

The plan of the paper is the following: 
In \secref{sec: Ginzburg-Landau in finite systems} we firstly recap canonical microscopic derivation of the Ginzburg-Landau model for non uniform FFLO superconductors.
Then we demonstrate  that the usual GL model is unbounded 
from below 
because it is unstable to formation of infinitely strong gradients near boundaries.
In \secref{sec: Ginzburg-Landau regularization} we construct a Ginzburg-Landau functional which has energy bounded from below by retaining higher order derivative terms in the expansion.
That allows us to in \secref{sec: 3D Superconductor transitions}, obtain the face, edge and vertex states without divergent energies. 
In \secref{sec: no divergence GL}  we construct an alternative form for the Ginzburg-Landau expansion which does not suffer from the divergence near the boundary, but does not capture the full set of different types of boundary states.

In \secref{sec: BdG approach} we solve numerically Bogoluibov-de-Gennes (BdG) equation to show that the states exist
in microscopic models that does not rely on any Ginzburg-Landau expansion. In the Appendix we confirm the existence of the boundary pair-density-wave by solving a more general BdG including Hartree terms.

\section{ Breakdown of the standard Ginzburg-Landau approach in finite systems} \label{sec: Ginzburg-Landau in finite systems}

\subsection{Ginzburg-Landau Model}
The Ginzburg-Landau description of superconductors in the presence of fermionic population imbalance was derived from microscopic theory
for superconductors in \cite{Buzdin1997}.
The free energy functional reads $F[\psi] = \integral{\Omega}{} \F \diffD{x}{d}$ where the free energy density $\F$ is
\begin{equation} \label{eq: Free energy functional, original}
\begin{aligned}
\F =   & \alpha |\psi|^2 + \beta |\grad{\psi}|^2 + \gamma |\psi|^4 + \delta |\laplace{\psi}|^2 + \\
& \mu |\psi|^2 |\grad{\psi}|^2 + \frac{\mu}{8} \big[ (\psi ^*\grad{\psi})^2 + \cc \big] + \nu |\psi|^6 ,
\end{aligned}
\end{equation}
where $\psi$ is a complex field referred to as the superconducting order parameter and $\cc$ denotes complex conjugation. The coefficients $\alpha, \gamma$, and $\nu$ depend on the fermionic population imbalance $H$ and temperature $T$ accordingly \cite{Buzdin1997}
\begin{align} 
\alpha & = -\pi N(0) \left( \frac{1}{\pi} \ln\frac{T_c}{T} + K_1(H,T) - K_1(0,T_c)\right), \label{eq: alpha coefficient} \\
\gamma  & = \frac{\pi N(0) K_3 (H,T)}{4}, \label{eq: gamma coefficient} \\
\nu & = -\frac{\pi N(0)K_5(H,T)}{8}, \label{eq: nu coefficient}
\end{align}
where $N(0)$ is the electron density of states at the Fermi surface, $T_c$ is the critical temperature at zero $H$ and we have defined the functions
\begin{equation} \label{eq: K-functions}
K_n (H,T)= \frac{2T}{(2 \pi T)^n}  \frac{(-1)^n}{(n-1)!} \Re \left[  \Psi ^{(n-1)}(z)\right],
\end{equation}
where $z = \frac{1}{2}- \ii \frac{H}{2 \pi T}$ and $ \Psi ^{(n)}$ is the polygamma function of order $n$.
The remaining coefficients are given as $\beta = \hat{\beta} v_\Fermi ^2 \gamma$, $\delta = \hat{\delta} v_\Fermi^4 \nu$, and $\mu = \hat{\mu} v_\Fermi^2 \nu$, where $v_\Fermi$ is the Fermi velocity and $\hat{\beta}, \hat{\delta},\hat{\mu}$ are positive constants that depend on the dimensionality $d$. In one dimension we have $\hat{\beta}=1$, $\hat{\delta}=1/2$, and $\hat{\mu}=4$, in two dimensions we have $\hat{\beta}=1/2$, $\hat{\delta}=3/16$, and $\hat{\mu}=2$, and in three dimensions we have $\hat{\beta}=1/3$, $\hat{\delta}=1/10$ and $\hat{\mu}=4/3$. The Ginzburg-Landau description is valid in the parameter regime in which the highest order terms are positive, that is where $\nu$ is positive. In the parameter regime in which $\beta$ is negative, inhomogeneous order parameters may be energetically favorable.
The first considered structures of the order parameter are the so-called Fulde-Ferrell (FF) state $\psi_\FF \propto \e ^{\ii px}$ and the Larkin-Ovchinnikov (LO) state $\psi_\LO \propto \cos px$, with $p^2 = -\frac{\beta}{2 \delta}$ and transition into the normal state at $\alpha = \alpha_c^{\bulk} = \frac{\beta ^2}{4 \delta}$.

Inhomogeneous states can appear when the term $| \nabla \psi|^2$ has a negative prefactor. In this parameter regime, it is necessary to include higher order terms, resulting in a free energy density expansion in \eqref{eq: Free energy functional, original}.
Where term $| \nabla \psi|^2$ favors creation of the gradients, while the positive term $|\Delta \psi |^2$ is added to bound gradients from above.
However we will show that in some cases, the inclusion of the stabilizing term $|\Delta \psi |^2$ is not sufficient. In small systems and generically in finite systems in two dimensions or higher, there exist solutions that satisfy $\Delta \psi = 0$. These states are often characterized by enhancement of $\psi$ close to the boundary. The associated energy of such state can potentially diverge and one needs to resort to a more general Ginzburg-Landau theory.

\subsection{Small systems} \label{subsec: Small systems}

Consider first a one dimensional domain $\Omega = [-L/2, L/2]$ of length $L$. The equation $\Delta \psi = 0$ is satisfied by the real field $\psi (x) = q x$, for any parameter $q$. The modulus $|\psi|$ increases linearly as the boundaries are approached, resulting in increasing potential energy density closer to the boundaries. As system size increases, the penalizing potential energy is non-negligible, making the linear solution less energetically beneficial. Therefore, the linear solution is not energetically preferable over conventional one dimensional inhomogeneous structures, such as $ \psi_\LO \propto \cos p x$ or $\psi_\FF \propto e^{\ii p x}$ if $L \gtrsim 1/p$, where $p^2 = -\frac{\beta}{2\delta}$.
However, for small system sizes, the potential term does become  negligible in comparison to the beneficial gradient term $\beta |\nabla \psi| ^2 = \beta q^2$. To lowest order in $L$, minimizing with respect to $q^2$ gives
\begin{equation}
q^2 = -\frac{2(\alpha L^2 + 12\beta)  }{5\mu L^2}, \qquad F = - \frac{(\alpha L^2 + 12 \beta )^2}{60 \mu L},
\end{equation}
with transition into the normal state at $\alpha = \alpha_{c1} = -12 \beta /L^2$. In the limit of infinitesimal system size $L \to 0$, we find that $\alpha_{c1}\to \infty$, and the associated momentum $q$ and energy $F$ diverges. Note that the value of $\psi $ at the boundary, $\psi (\pm L/2) = \pm qL/2$, does not vanish in this limit since $q \propto 1/L$. Consequently the value of $\psi$ at the boundary is independent of system size for sufficiently small systems.

\begin{figure}[H]
\center
\includegraphics[width=8cm]{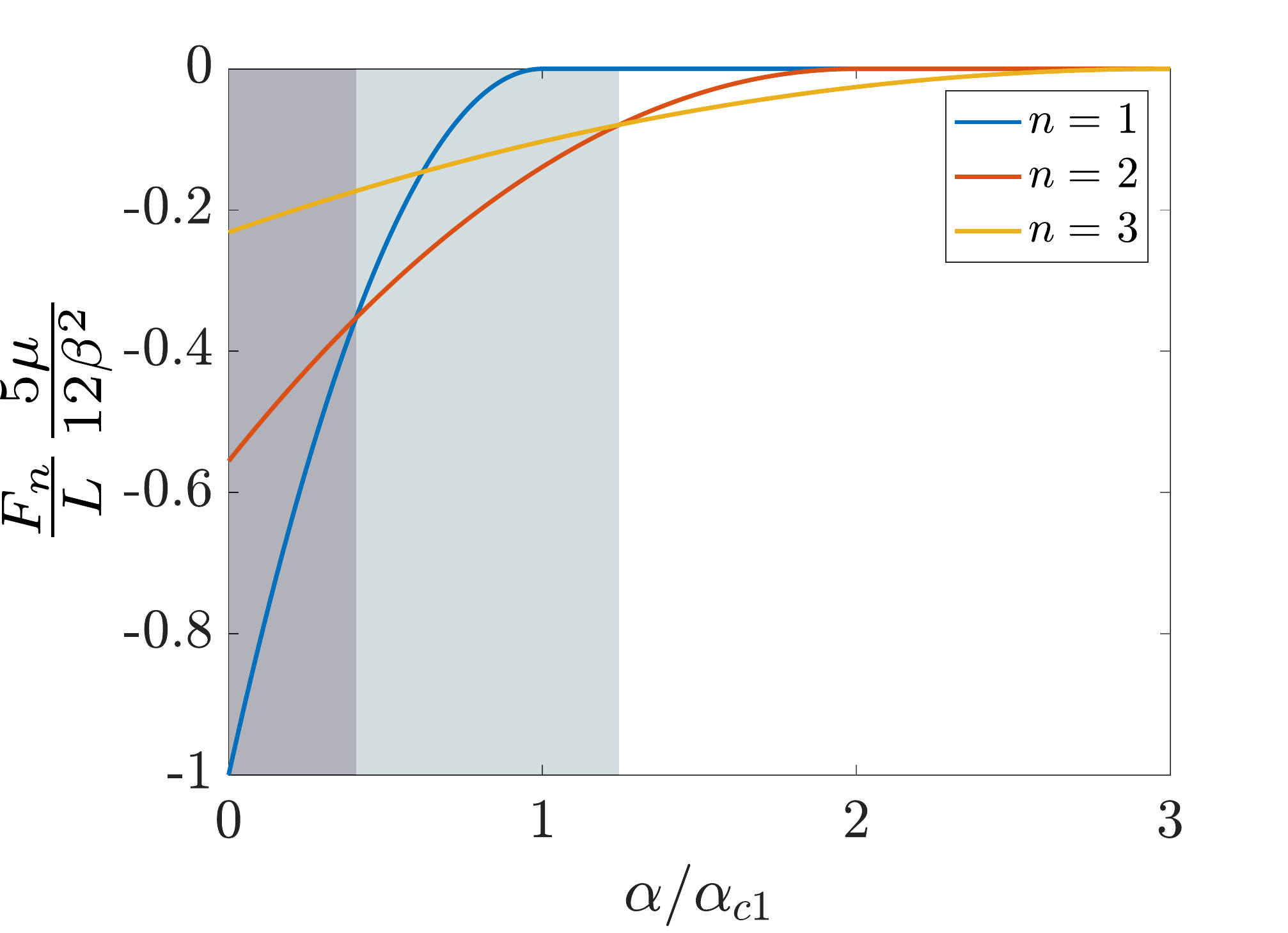}
\caption{
Free energies associated with the multi-linear solution $\psi_n (\vec{x}) = q_n \prod_{j=1}^n x_j$ in three dimensions, in small system sizes. As $\alpha$ is increased, we transition from the linear solution $n=1$, to the bilinear solution $n=2$ and finally the tri-linear solution $n=3$ before entering the normal state.} \label{fig: Dimensions multi-linear solution}
\end{figure}

The effect  generalizes to systems in higher dimensions. Consider the $d$-dimensional cube with volume $L^d$. The linear solution here generalizes to a multi-linear solution $\psi_n (\vec{x}) = q_n \prod_{j=1}^n x_j$, where $n\leq d$. Analogously to the one-dimensional case we find
\begin{equation}
\begin{aligned}
q^2_n &  = - \frac{1}{n}\left[  \frac{5}{3} \left( \frac{2}{L} \right)^2\right]^{n-1} \frac{2(\alpha L^2 + 12n\beta)}{5 \mu  L^2}, \\
F_n & = -L^{d-1}  \frac{1}{n}\left( \frac{5}{18} \right)^{n-1} \frac{(\alpha L^2 + 12 n \beta )^2}{60 \mu L},
\end{aligned}
\end{equation} 
with transition into the normal state at $\alpha_{cn} = n \alpha_{c1}$. Consequently we expect that in higher dimensions, we will see a sequence of transitions from $(n=1) \to (n=2) \to \ldots \to (n=d)$ before transitioning into the normal state, as seen in \figref{fig: Dimensions multi-linear solution} in three dimensions. Studying the energy $F_n$, we see that the free energy would approach some negative constant in two dimensions, and zero from below in three dimensions, as the system size approaches zero. Similarly to the one-dimensional case, the multi-linear solution $\psi_n$ is constant and independent of system size at the vertices of the $n$-dimensional cube, for significantly small systems.

\subsection{Boundary state instability in Ginzburg-Landau model}
\begin{figure}[H]
\center
\includegraphics[width=8cm]{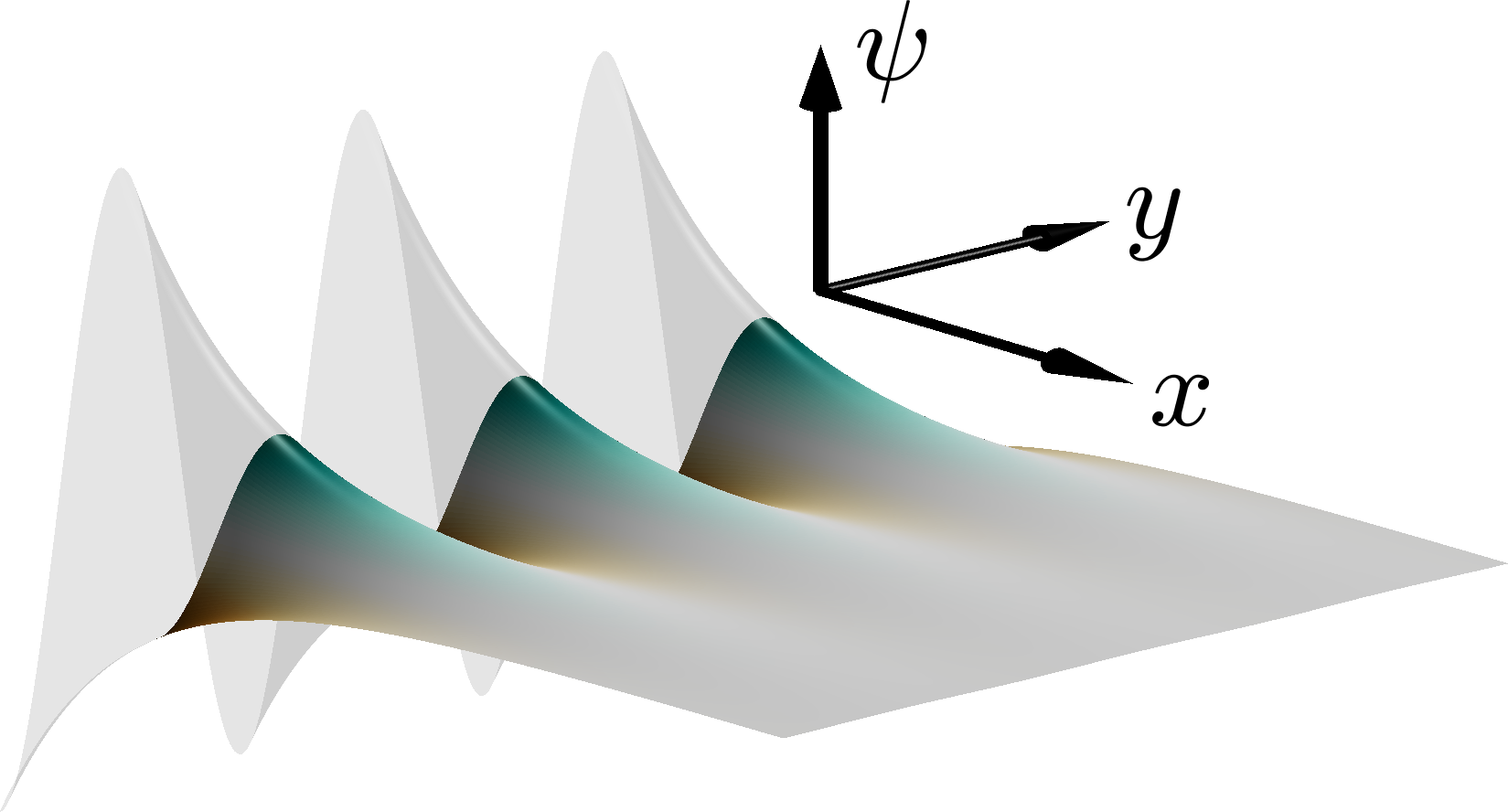}
\caption{
Order parameter of two-dimensional boundary state with density modulation along the boundary. In an infinite system, such a state would not be favorable due to divergent modulus in one direction. However, in the presence of boundaries, the segment of divergent modulus can be cut off, depicted on the image by the transparent surface.} \label{fig: example 2D surface state}
\end{figure}

Consider a two-dimensional domain $\Omega$ in the presence of boundaries. We can represent the domain  $\Omega$ as section of the complex plane, by defining a complex coordinate $\zeta=x+\ii y$. The order parameter $\psi$ is now a function of $\zeta$. Clearly if  $\psi$ is an analytic complex valued function (i.e. it satisfies the Cauchy-Riemann equations), it follows automatically that $\Delta \psi = 0$. The system becomes unstable towards states of the form $\psi (\zeta) =  A e^{k \zeta}$ with $|k| \to \infty$, associated with a negative divergent energy. Crucially, the formations of these states require the existence of a boundary. If the complete two-dimensional plane were considered, the solution  $\psi (\zeta) = A e^{k \zeta}$ would diverge in some direction, which would be penalized by the potential terms. However, the divergence of $|\psi|$ can be cut off using the boundary of the system, as seen in \figref{fig: example 2D surface state}.

For example, consider the half-infinite two-dimensional domain $x \geq 0$. An analytic function which satisfies $\Delta \psi = 0$ is
\begin{equation}\label{analyt_fun}
\psi = A e^{-kx+\ii k y}
\end{equation}

with $k>0$. This state is characterized by phase-wave modulation (FF state) tangential to the boundary and exponential decay perpendicular to the boundary. We compute the energy density along the boundary (that is in the $y$-direction) by integrating over $x$ and find
\begin{equation} \label{eq: Divergent 2D energy}
\mathcal{F}_\parallel = \frac{6 \alpha |A|^2 + 3\gamma |A|^4 + 2\nu |A|^6}{2k} + \frac{2 \beta |A|^2 + \mu |A|^4}{2}k.
\end{equation}
We see that as long as the amplitude $|A|^2 < -2 \beta / \mu$, the energy diverges to $-\infty$ as $k$ increases. This shows that the model has an instability towards formation of boundary states with rapid modulation along the boundary. Analogous calculation for the state
\begin{equation}
\psi = A e^{-kx} \cos ky
\end{equation}

would yield the same conclusion, see \figref{fig: example 2D surface state}.

\section{Generalisation of the  Ginzburg-Landau model} \label{sec: Ginzburg-Landau regularization}
In contrast to the example of linear solutions in small systems, see \secref{subsec: Small systems}, the two-dimensional solutions with modulation along the boundary and exponential decay from the boundary exist in any system in the presence of boundaries, regardless of the system size. The divergence in energy and Cooper pair momentum is indicative towards the existence of superconducting boundary states.
Note, however that GL solutions can only represent the long-wave-length physics of how order parameter decays away from the $1 / k_F$ length scales from the boundary.
Hence, in order to find a suitable theory which is bounded from below, we have to include additional terms in the Ginzburg-Landau expansion.

We would like to find the lowest order term in both momentum and amplitude of the order parameter, which is sufficient to include in order to stabilize the theory. 
The origin of the inhomogeneous state is the negative term $|\grad{\psi}|^2$, which is of second order in both momentum and amplitude. Conventionally it is stabilized by the term $|\laplace{\psi}|^2$, which is of fourth order in momentum and second order in amplitude.
Since there exists states that satisfy $\Delta\psi = 0$, we have to resort to higher order in either momentum or amplitude. Terms such as $|\grad{\laplace{\psi}}|^2$, which is of sixth order in momentum and second order in amplitude, would not be helpful since it is automatically zero if $\Delta \psi = 0$. Therefore we resort to retaining the term which is of fourth order in momentum and fourth order in amplitude, which reads
\begin{equation}\label{eq: Regularizing term}
	\kappa|\nabla\psi|^4.
\end{equation}
Since this term is of higher order in amplitude than the lowest order beneficial gradient term, there will exist inhomogeneous superconducting states for any $\alpha$, as long as the gradient coefficients are negative. 
This term was used to find two-dimensional solutions in  \cite{barkman2018surface}.
Here we  derive from the  microscopic theory the following estimate for the coefficient $\kappa$ in \eqref{eq: Regularizing term}
\begin{equation}
\kappa \simeq \left(-\frac{29}{32}\right) \frac{\pi   N(0) v_\Fermi^4 \Omega_d}{2} K_7(H,T) ,
\end{equation}
where factor $\Omega_d$ depends on the dimension $d$, where $\Omega_1 = 1$, $\Omega_2 = 3/ 8$ and $\Omega_3 = 1/5$. Studying the functions $K_5(H,T)$ and $K_7 (H,T)$, we see that there exist an overlapping region in which both functions are negative, which implies that the additional term is positive simultaneously as the previously higher order terms. In order to study a minimal model, we proceed with only retaining the term in \eqref{eq: Regularizing term} in the regularized free energy expansion, even though additional terms also proportional to $K_7 (H,T)$ could be included. For example, in principle we could have included a potential term proportional to $|\psi|^8$. However, since the sixth order potential term remains non-zero and positive, the inclusion of this term is not so important. 

Let us now consider the previous example of the boundary state $\psi = A e^{-kx + \ii ky}$, which previously was associated with negative infinite energy and infinite momentum $k$. With the inclusion of the term in \eqref{eq: Regularizing term}, the associated free energy in \eqref{eq: Divergent 2D energy} obtains the additional term
\begin{equation} \label{eq: Additional term in energy example}
\kappa |A|^4 k^3,
\end{equation}
which makes the free energy bounded from below and the optimal value of $k$ is now finite. 

\section{Sequential phase transitions in three dimensions} \label{sec: 3D Superconductor transitions}

Having obtained the Ginzburg-Landau free energy expansion without the spurious divergences
enables us to study higher dimensional systems. We will numerically minimize the free energy functional using the nonlinear conjugate gradient method, parallelized on the CUDA-enabled NVIDIA graphical processing unit (GPU). It is convenient to introduce rescaled coordinates and parameters defined accordingly: $\tilde{\psi} = \psi / |\psi_\U |$, $\tilde{\alpha} = \alpha /  \alpha_\U $, $\tilde{x} = p x $, where $|\psi_\U|^2 = - \frac{\gamma}{2\nu}$, $\alpha_\U = \frac{\gamma^2}{4 \nu}$ and $p^2 = -\frac{\beta}{2\delta}$. Expressed in these quantities, the free energy reads $F[\psi] =  \frac{\alpha_\U |\psi_\U|^2}{p^d} \tilde{F}[\tilde{\psi}]$, where $\tilde{F}[\tilde{\psi}] = \int _\Omega \tilde{\mathcal{F}} \diffD{\tilde{x}}{d}$, where the rescaled free energy density $\tilde{\mathcal{F}}$ is identical to \eqref{eq: Free energy functional, original}, but the coefficients have been replaced accordingly: $\alpha \mapsto \tilde{\alpha}$, $\beta \mapsto \tilde{\beta}$, and so on, where $\tilde{\gamma}=-2\tilde{\nu}=-2$, $\tilde{\beta}= -2 \tilde{\delta} = -2 \hat{\beta}^2 / \hat{\delta}$, and $\tilde{\mu} = \hat{\beta} \hat{\mu} / \hat{\delta}$. Among these coefficients, all are constant except $\tilde{\alpha}$, which parametrizes both temperature $T$ and fermionic population imbalance $H$. With the inclusion of the additional term in \eqref{eq: Regularizing term}, in rescaled coordinates, the coefficient $\tilde{\kappa}$ reads
\begin{equation} \label{eq: rescaled regularized coefficient}
\tilde{\kappa} =- \frac{1}{v_\Fermi ^4} \frac{\hat{\beta}^2}{2 \hat{\delta}^2} \frac{\kappa \gamma}{\nu^2}.
\end{equation}

Specifically in three dimensions we have that $\tilde{\kappa} = \frac{145 K_3 K_7}{18 K_5^2}$, which is not constant in the rescaled units. However, studying the functions $K_n$ in \eqref{eq: K-functions}, we see that $K_n \propto \frac{T}{T^n} \cdot f_n(H/T)$, where $f_n$ is some elementary function. Therefore, if we study a line in the $TH$-plane where $H/T$ is constant, the rescaled parameter $\tilde{\kappa}$ will also be constant. Particularly, we will study the line where $H/T = 2\pi / 3$, along which $\tilde{\kappa} \simeq 0.5752$. On a technical note, since we are working in rescaled units, where we measure length in units of $p^{-1} \propto \sqrt{K_5 / K_3}$, we also have to alter the rescaled length accordingly in order to appropriately describe the sample of fixed size.

\begin{figure}[H]
	\centering
	\includegraphics[width=0.99\linewidth]{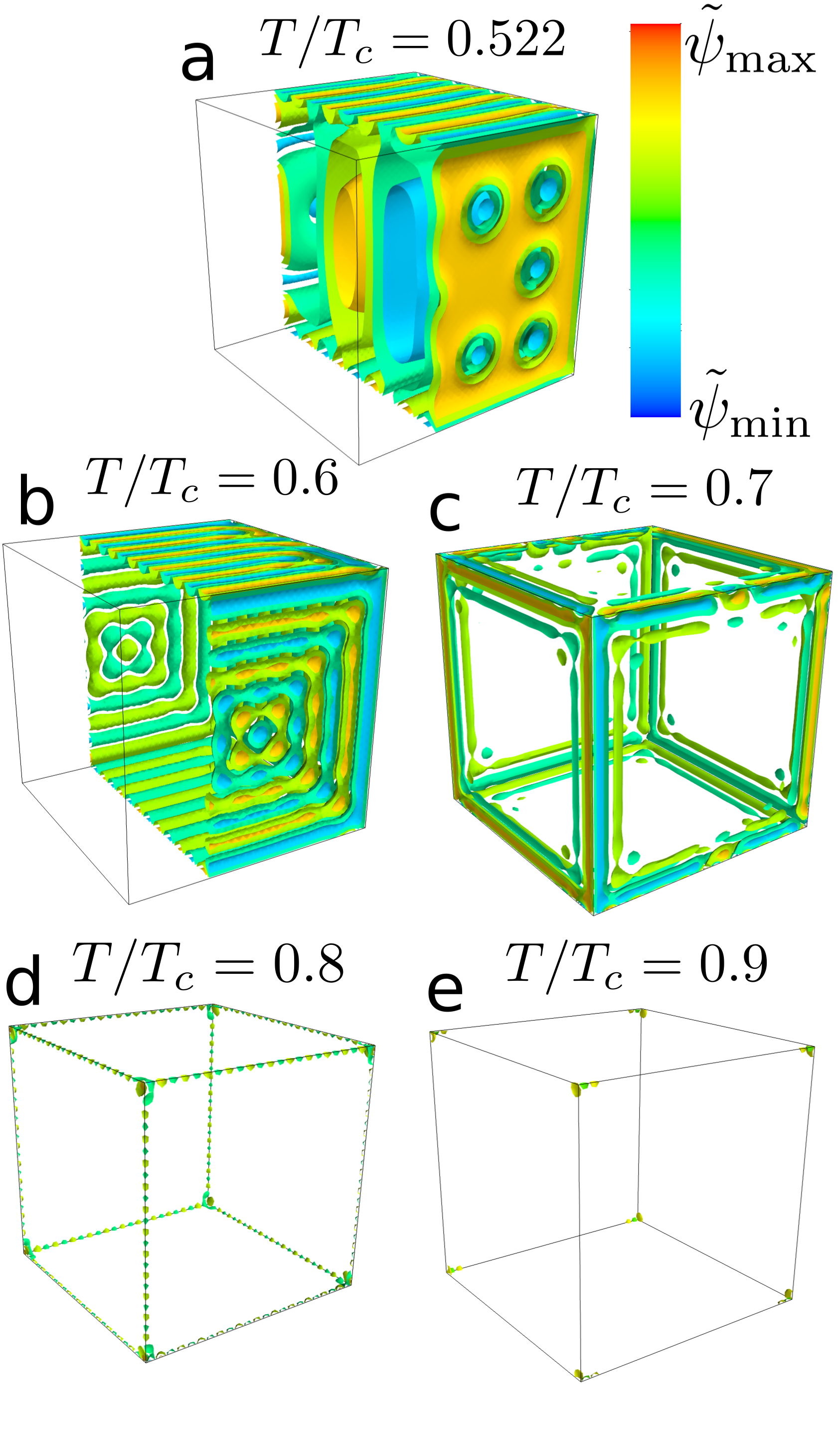}
	\caption{
Pair-density-wave states in a three-dimensional cube, of size significantly bigger than characteristic length of the variation of order parameter, obtained from Ginzburg-Landau theory, for various temperatures $T$ and fermionic population imbalances $H$. As temperature is increased, we observe a sequence of phase transitions. For low temperatures, superconductivity exists in the whole system. In the interior the solution is of Larkin-Ovchinnikov type
but has higher gradients and different pattern on the boundary due to reasons explained in the text. As temperature is increased, superconductivity vanishes first in the bulk, secondly on the faces and thirdly on the edges, such that eventually superconducting gap only exists in the vertices of the cube. The fermionic population imbalance $H = 2\pi T/3 $ for each of the illustrations (\textbf{a}-\textbf{e}). The minimal  and maximal values $\tilde{\psi}_\min$ and $\tilde{\psi}_\max$ are different for each of the different illustrations. In \textbf{a} $-\tilde{\psi}_\min = \tilde{\psi}_\max = 0.7$. In \textbf{b} $-\tilde{\psi}_\min = \tilde{\psi}_\max = 0.4$. In \textbf{c} and \textbf{d} $-\tilde{\psi}_\min = \tilde{\psi}_\max = 0.2$. In \textbf{e} $-\tilde{\psi}_\min = 2\tilde{\psi}_\max = 0.08$.
}\label{fig: 3D Ginzburg-Landau simulations}
\end{figure}

The obtained solutions are shown in  \figref{fig: 3D Ginzburg-Landau simulations}. We see that as temperature is increased, there are multiple phase transitions.
The dimensionality of the transition  sequentially decreases. At the lowest temperature, the system is in a superconducting state with a non-uniform order parameter in the bulk. Note that deep in the bulk the solution is of Larkin-Ovchinnikov type.
On the boundary, due to discussed above reasons we expect larger gradients. This is indeed clearly seen on the upper panel of Fig. \ref{fig: 3D Ginzburg-Landau simulations}.
Note, that the order parameter configurations on the faces are very different from the Larkin-Ovchinnikov solutions in the bulk.
 
As temperature is increased, we first transition into a superconducting state which allows for modulation on the faces of the cube, while the bulk now has transition into a normal state. Second, when the temperature is elevated further,  the faces become normal too, but  superconductivity  survives on the edges of the cube. Thirdly, the edges become normal and only the vertices remain superconducting, before finally the fully normal state is entered at an even higher temperature. 

\section{Resolving the boundary instability in an alternative Ginzburg-Landau expansion} \label{sec: no divergence GL}

Let us consider how  to obtain a model which does not suffer from divergences  near a boundary without introduction of the higher order terms, as in \secref{sec: Ginzburg-Landau regularization}.

Firstly, observe  that when one uses the usual microscopic derivation of the  model \eqref{eq: Free energy functional, original} one does not obtain fully unambiguously the form of the gradient terms.
The ambiguity is due to freedom in integration by parts.
For example, the term $-\psi \Delta \psi^* + c.c.$ is  not discriminated from $|\nabla\psi|^2$.
Other way to put it is that the standard microscopic derivation fixes Euler-Lagrange equations but does not fully fix the boundary conditions.

The GL free energy in \eqref{eq: Free energy functional, original} is unbounded from below due to the existence of boundary states satisfying $\Delta \psi = 0$.
We can rewrite the Laplacian term $|\Delta \psi|^2$ using integration by parts as
\begin{equation}\label{lap2}
\begin{aligned}
\int_\Omega |\Delta \psi|^2 \diff{V}  = \int_\Omega \sum_{i, j = 1}^{3} |\partial_i \partial_j \psi|^2 \diff{V} + \oint_{\partial \Omega} f \diff{S}, \\
f =\sum_{i, j = 1}^{3} \Big( \big( \partial_i^2 \psi^*\big) \partial_j \psi - \big( \partial_i \partial_j \psi^*\big) \partial_i \psi \Big) n_j,
\end{aligned}
\end{equation}

where $\vec{n}$ is the normal to the boundary $\partial \Omega$. Since the standard microscopic derivation does not unambiguously determine the boundary contribution, we consider now the free energy expansion where $|\Delta \psi|^2 $ is replaced by $\sum_{ij} |\partial_i \partial_j \psi |^2$ and the boundary term $f$ is discarded.
The boundary state in \eqref{analyt_fun} does not nullify $\sum_{ij} |\partial_i \partial_j \psi |^2$, and only linear function $\psi = a + \sum_{i = 1}^{3} b_i x_i$ with constants $a,\ b_i$ does.
Note that in this alternative form, the free energy does not suffer from boundary divergences and the inclusion of higher order terms is unnecessary.

\begin{figure}[H]
	\centering
	\includegraphics[width=0.99\linewidth]{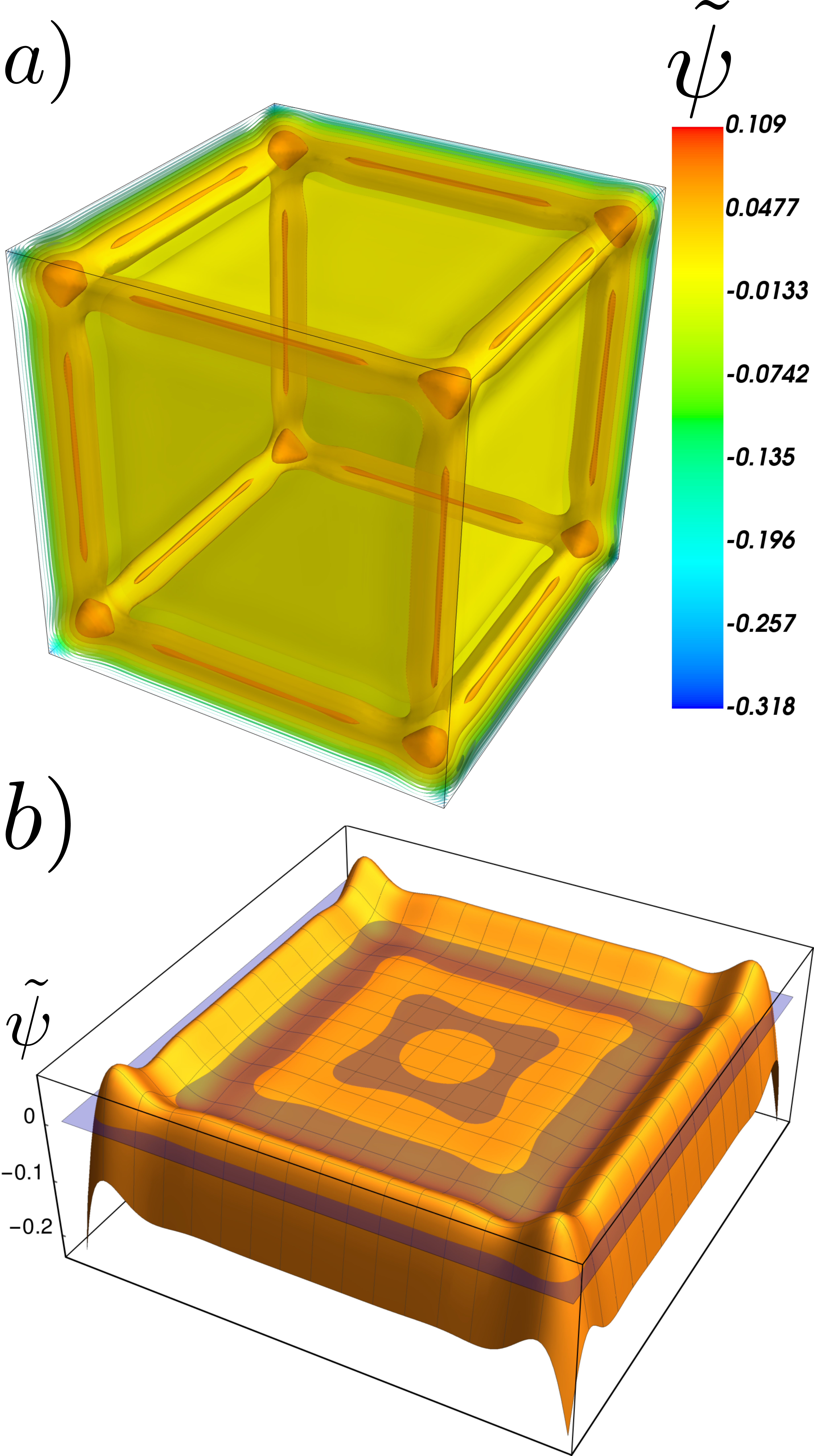}
	\caption{
		Boundary pair-density-wave state in a three-dimensional cube obtained in rescaled Ginzburg-Landau model \eqref{eq: Free energy functional, original} with Laplacian term replaced by $\sum_{ij} |\partial_i \partial_j \psi |^2$ at $\tilde{\alpha} = 4$.
		The sample is above bulk critical temperature but below the boundary critical temperature.
		Order parameter has zero imaginary part and real part is presented on panels:
		\textbf{a)} Isosurfaces of $\psi$ in a three dimensional sample that was simulated numerically.
		Observe that it is almost zero in the bulk and non-zero on the boundary (most prominent in vertices).
		\textbf{b)} (orange) Order parameter in horizontal cross section of the three dimensional sample a) at the middle of the system.
		(blue) zero of $\psi$.
		In this approximation even though order parameter is clearly bigger in the vertices and edges that on faces, there is no clearly identifiable sequence of four phase transitions as in the case presented on \figref{fig: 3D Ginzburg-Landau simulations}.
		Instead the model has only three phases: bulk and boundary pair-density-wave and the normal phase.
	}\label{fig: new Laplacian regularization}
\end{figure}

We simulated numerically the GL model given by \eqref{eq: Free energy functional, original} with Laplacian term replaced by $\sum_{ij} |\partial_i \partial_j \psi |^2$.
The typical configuration of the order parameter is presented in \figref{fig: new Laplacian regularization}.
The main difference from the GL model with usual Laplacian term and higher order terms, see \figref{fig: 3D Ginzburg-Landau simulations}, is that we do not recover the full four-step sequence (bulk-faces-edges-vertices) of the phase transitions in the cuboid geometry.
Instead in this approximation it retains a two-step transition only from bulk to boundary superconductivity and then to normal state everywhere.

Now we can construct an analytical guess for the edge state, which is the same as in \cite{barkman2018surface}.
Namely, let us consider superconductor positioned at $x,\ y > 0$ and for all $z$.
Then configuration of $\psi$ in \figref{fig: new Laplacian regularization}b can be approximated by
\begin{equation}
\psi(x, y) = \eta(x) + \eta(y),\quad \eta(x) = e^{-k x} \cos(q x + \phi).
\end{equation}

Even though this configuration has enhanced $\psi$ on the edges, due to the overlap of $\eta(x)$ and $\eta(y)$, 
it will, in this approximation, have the same critical temperature, as the critical temperature of the face superconductivity in the limit of infinitely large sample.

\section{Microscopic Bogoluibov-de Gennes (BdG) Approach} \label{sec: BdG approach}
\subsection{Derivation of BdG from path integral}
We start from Fermi-Hubbard model in path integral formulation defined by action (first part of the derivation follows \cite{altlandsimonsBook})
\begin{equation}\label{FermiHubbard}
\begin{aligned}
S = \int_0^{\frac{1}{T}} d\tau \sum_{i, j = 1, \sigma = \downarrow}^{i, j = N, \sigma =\uparrow} \psi^\dagger_{i \sigma} (\partial_\tau + h_{i j \sigma}) \psi_{j \sigma}\\
- V \sum_{i = 1}^{N} \psi^\dagger_{i \uparrow} \psi^\dagger_{i \downarrow} \psi_{i \downarrow} \psi_{i \uparrow}
\end{aligned}
\end{equation}

with Grassman fields $\psi_{i \sigma}(\tau), \psi^\dagger_{i \sigma}(\tau)$ corresponding to fermionic creation and annihilation operators defined on some lattice indexed by $i, j \in [1, N]$.
Kinetic part of action is given by $h_{i j \sigma} = - (\mu + \sigma h) \delta_{i, j} - t \delta_{|i - j|, 1}$, with chemical potential $\mu $, spin-population imbalance $h$ and nearest-neighbor hopping parameter $t$.
Interaction is set by on-site attractive potential $V > 0$, which leads to $s$-wave pairing.
Partition function is given by path integral
\begin{equation}
Z = \int D[\psi^\dagger, \psi] e^{-S}.
\end{equation}

We perform Hubbard-Stratonovich transformation, which introduces auxiliary bosonic field $\Delta_i(\tau)$. Up to a constant the interaction term becomes:
\begin{equation}
\begin{aligned}
\exp\left[ V \int d\tau \sum_{i} \psi^\dagger_{i \uparrow} \psi^\dagger_{i \downarrow} \psi_{i \downarrow} \psi_{i \uparrow} \right]
= \int D[\Delta^\dagger, \Delta]\\ \exp\left[ - \int d\tau \sum_{i} \left( \frac{\Delta^\dagger_i \Delta_i}{V} + \Delta^\dagger_i \psi_{i \downarrow} \psi_{i \uparrow} + \Delta_i \psi^\dagger_{i \uparrow} \psi^\dagger_{i \downarrow} \right) \right].
\end{aligned}
\end{equation}

By introducing Nambu spinors
\begin{equation}
\Psi^\dagger_i = \left( \psi^\dagger_{i \uparrow}, \psi_{i \downarrow} \right),\ \ \ 
\Psi_i = 
\begin{pmatrix}
\psi_{i \uparrow}\\
\psi^\dagger_{i \downarrow}
\end{pmatrix}
\end{equation}

the action becomes
\begin{equation}
S = \int d\tau \sum_i \frac{\Delta^\dagger_i \Delta_i}{V} + \sum_{i j} \Psi^\dagger_i \left( \delta_{i j} \partial_\tau + H_{i j} \right) \Psi_j
\end{equation}

With $2 N \times 2 N$ matrix $H$ defined by:
\begin{equation}\label{eq: Mean-field Hamiltonian}
H_{i j} = 
\begin{pmatrix}
h_{i j \uparrow} & \delta_{i j} \Delta_i \\
\delta_{i j} \Delta^\dagger_i & - h_{j i \downarrow}
\end{pmatrix}.
\end{equation}

Integrating out fermionic degrees of freedom we obtain
\begin{equation}
Z = \int D[\Delta^\dagger, \Delta] \exp\left[ \ln{\det\left( \partial_\tau + H \right)} - \int d\tau \sum_{i} \frac{\Delta^\dagger_i \Delta_i}{V} \right].
\end{equation}

Note, that up till this point this model exactly corresponds to Fermi-Hubbard model \eqref{FermiHubbard}. To simplify it and proceed with mean field approximation we make two assumptions: 1) $\Delta_i$ is classical, i.e. it does not depend on $\tau$ 2) $\Delta_i$ does not fluctuate thermally, i.e. it extremizes the action $S$. 
Note that this mean-field nature of BdG approach will give us a sequence of second order transitions
for the superconductivity of the bulk, face, edge, and vertex states. Beyond mean-field approximation
the phase fluctuations make the face phase transitions of Berezinskii-Kosterlitz-Thouless type.

Next, one can simplify the action by reducing the term
\begin{equation}
\ln{\det\left( \partial_\tau + H \right)}
= \text{Tr} \sum_{n = -\infty}^\infty \ln\left( \ii \omega_n + H \right)
= - \text{Tr} \ln f(H)
\end{equation}
where last equality holds up to a constant, summation is performed over Matsubara frequencies $\omega_n = 2 \pi T \left( n + 1/2 \right)$ and Fermi factor $f(x) = 1/(e^{x/T} + 1)$. Note, that trace sums over $2 N$ diagonal elements of function of $2 N \times 2 N$ dimensional matrix $H$ and hence equals to just sum over a function evaluated at the eigenvalues of $H$. Finally, the action is given by
\begin{equation}\label{meanfieldS}
S = \sum_{i} \frac{|\Delta_i|^2}{T V} + \text{Tr}\ln f(H).
\end{equation}
Taking variation of \eqref{meanfieldS} with respect to $\Delta^*_i$ we obtain the self consistency equation\footnote{Note, that this formula is equivalent to the usual one with hyperbolic tangent since $- \vec{e_i}^T f(H) \vec{h_i} = \frac12 \vec{e_i}^T \tanh \frac{H}{2 T} \vec{h_i}$.} 
\begin{equation}\label{Delta_self_consistent} 
\Delta_i = - V \vec{e}_i^T f(H) \vec{h}_i = -V\langle c_{i\uparrow}c_{i\downarrow}\rangle
\end{equation}
where vectors, following \cite{nagai2012efficient}, are denoted as $(\vec{e}_i)_j = \delta_{i, j}$ and $(\vec{h}_i)_j = \delta_{i + N, j}$.
Note, that in this approach, based on path integral formulation, we obtain equation only for $\Delta_i$ \eqref{Delta_self_consistent}.
If one instead employs variational method self consistency equations are also obtained for Hartree terms, which shift the chemical potentials.
In the appendix we show that the boundary states, discussed here, exist in the model with Hartree term as well.

We solve the problem numerically by taking a recursive approach.
That is, we start with some initial guess for the pairing field $\Delta_i$.
Secondly, we diagonalize the Hamiltonian in \eqref{eq: Mean-field Hamiltonian}, i.e. we calculate its $2 N$ eigenvalues $E_n$ and eigenvectors $w_i^n = \begin{pmatrix} u_i^n \\ v_i^n\end{pmatrix}$.
Then $\Delta_i$ is updated through the self consistency equation \eqref{Delta_self_consistent}, which explicitly can be written as
\begin{equation}\label{eq: selfconsistent}
\Delta_i = - V \sum_{n = 1}^{2 N} u_i^n v_i^{n *} f(E_n)
\end{equation}

The process is repeated until convergence.

\subsection{Numerical results}
In this study we will investigate the surface properties, which requires significantly large system sizes. We consider both one- and two-dimensional systems, using open boundary conditions. In two dimensions we use a GPU-based approach, while in one dimension we use CPUs.

\begin{figure*}
\center
\includegraphics[height=6cm]{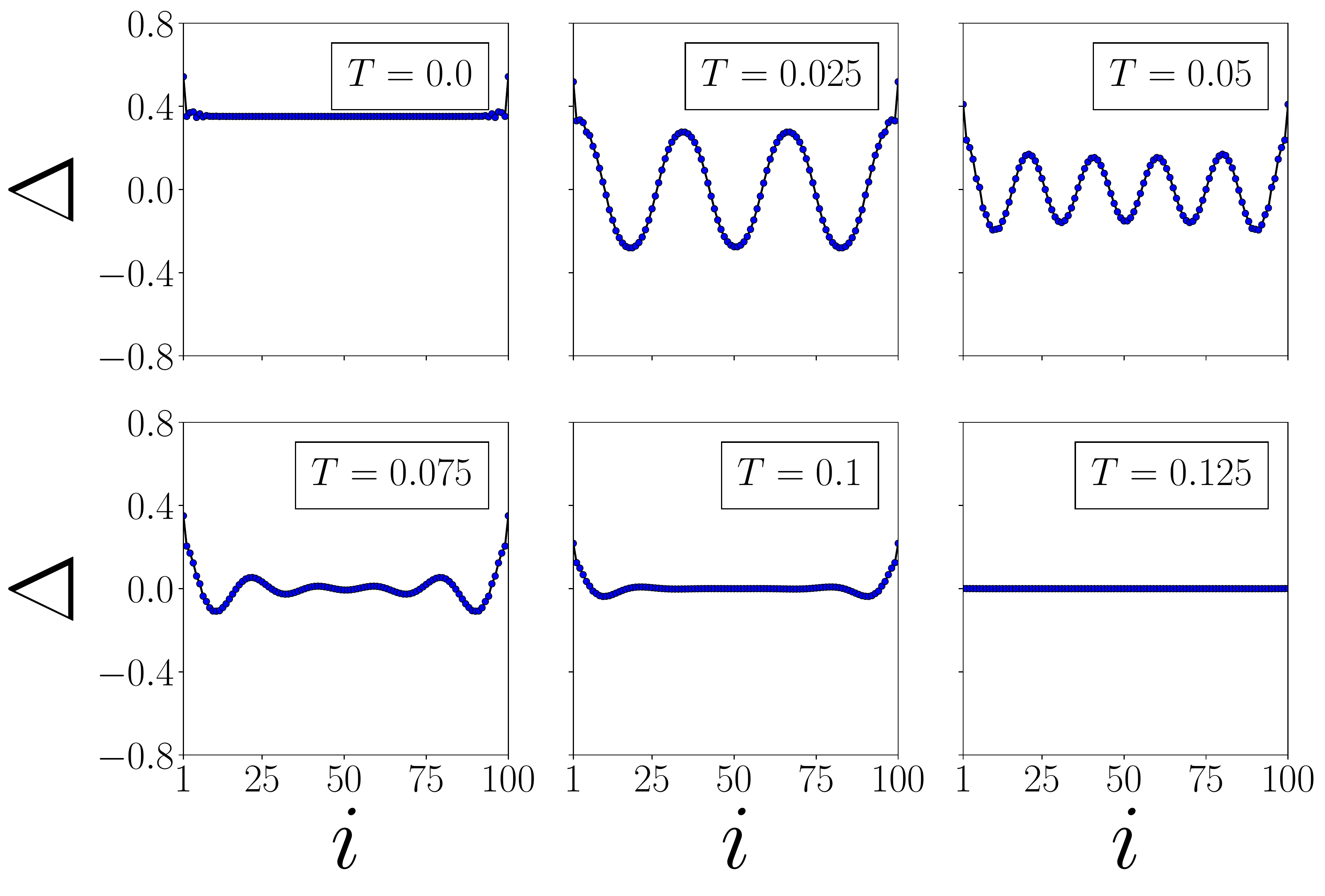}
\includegraphics[height=6cm]{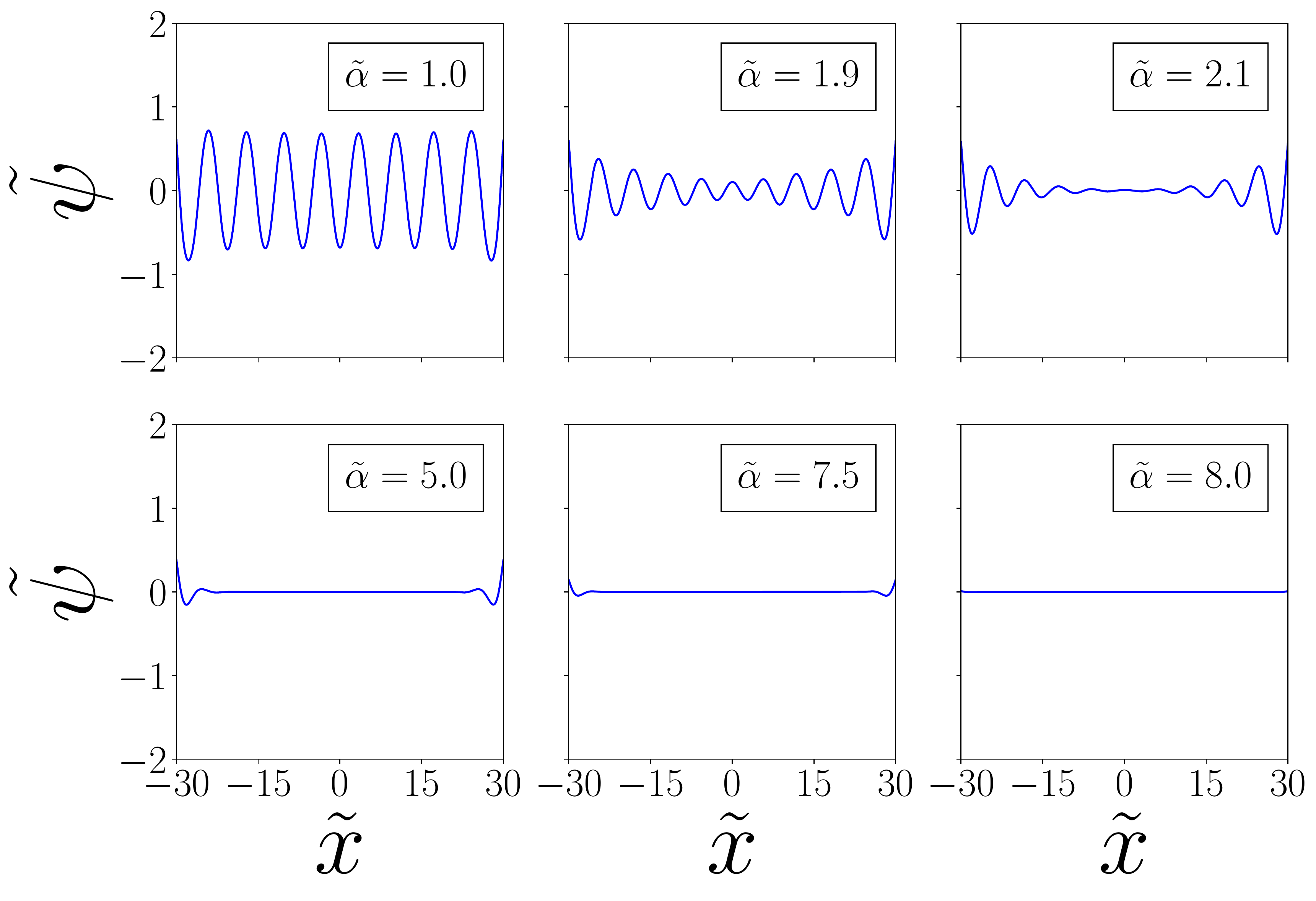}
\caption{
Left panels: Numerically calculated pairing field $\Delta$ by solving the BdG equation for various temperatures $T$, for some fixed non-zero spin-population imbalance $h=0.3$. As temperature is increased, we observe a sequence of phase transitions. Firstly, the system transitions from a uniform superconducting state with boundary enhancement, to the pair-density-wave FFLO state. Secondly, the pairing field vanishes in the bulk, while the pair-density-wave state remains non-zero close to the boundary, before finally transitioning fully into the normal conducting state. The behavior is of accordance with the prediction from Ginzburg-Landau theory in \cite{barkman2018surface}, from which the right panels have been taken. Remaining parameters are set to $t=1$, $V=2$ and $\mu = 0.5$.}\label{fig: 1D BdG}
\end{figure*}

The obtained order parameter for a one-dimensional system, while varying the temperature, is shown in  \figref{fig: 1D BdG}. The remaining parameters are set to $t=1$, $V=2$, $\mu = 0.5$ and $h=0.3$. Note that this parameter set does not correspond to the microscopic physics where Ginzburg-Landau expansion
was obtained, so we should not expect quantitative similarities. Rather the main question we ask if the boundary states
exist beyond the Ginzburg-Landau regime. To emphasize that the results do not require finetuning
in the Appendix we give an example of a boundary state in a different Bogoliubov-de-Gennes model with a Hartree term retained. We can notice, as temperature is increased, a transition from the uniform superconducting state with upshoot on the boundaries, into the non-uniform LO state and eventually the boundary state, referred to as the surface pair-density wave state, before finally transitioning into the normal state. The solutions are very similar to those obtain from the Ginzburg-Landau approach in \cite{barkman2018surface}, shown on the same figure. Note however that the microscopical considerations using BdG are not performed in the weak-coupling limit where Ginzburg-Landau theory applies. Therefore, a quantitative agreement between these two models are neither necessary nor trivially expected. We emphasize that the similarity between the two approaches concern the order parameter structure and not quantitative values in the phase diagram.

Note, that in the tight-binding model, non-zero values of the pairing field $\Delta_i$ on the boundary, as in \figref{fig: 1D BdG}, does not mean that it is discontinuous.
Instead $\Delta_i$ decays to zero outside the superconductor over length scale corresponding to the localization of the Wannier function’s tail or lattice spacing.

\begin{figure}[H]
\center
\includegraphics[width=8cm]{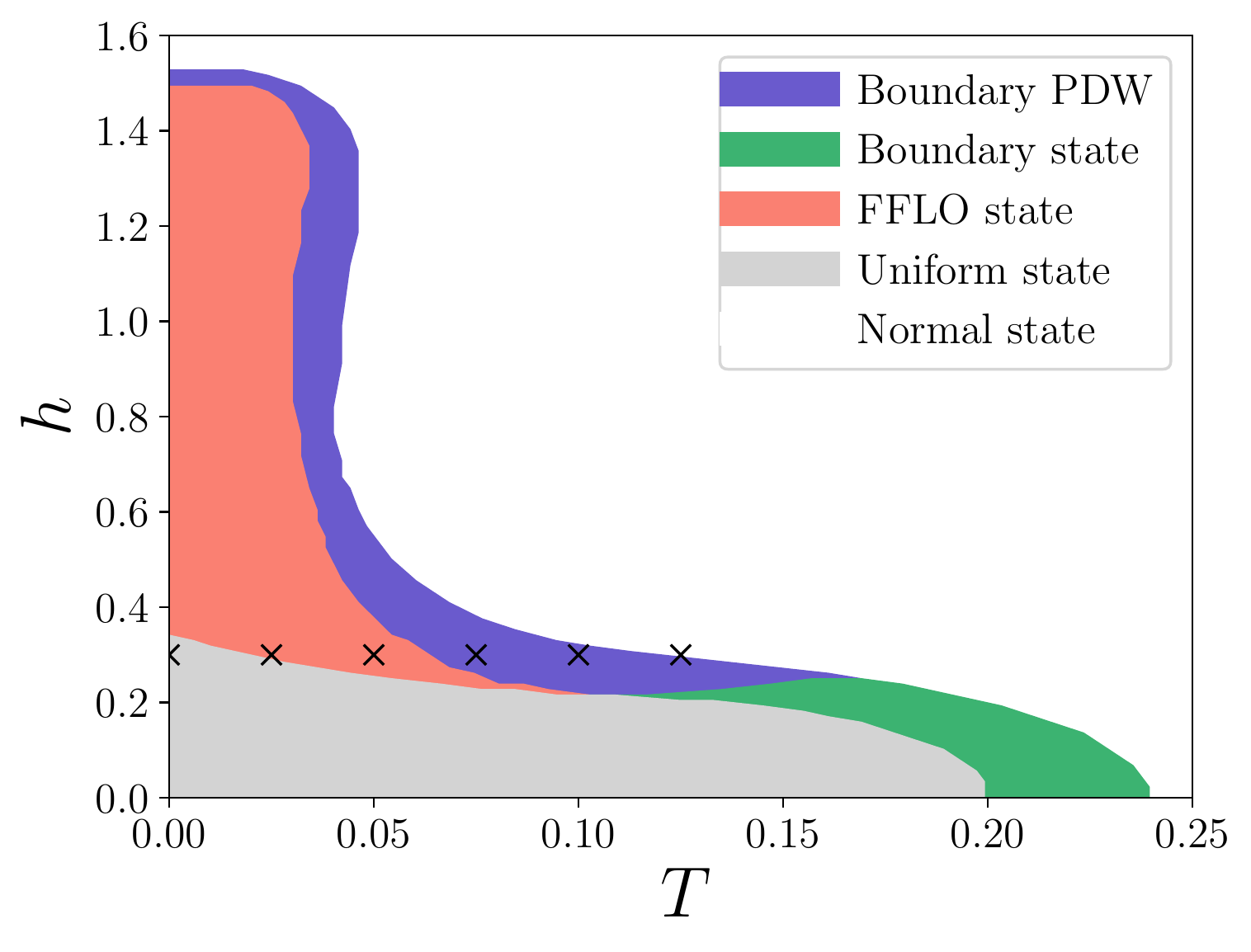}
\caption{
Phase diagram obtained from solving the BdG equation for various temperatures $T$ and spin-populations imbalances $h$, in one dimension in the model \eqref{meanfieldS}. We see the appearance of regimes in the phase diagram, in which superconductivity does not exists in the bulk, but exists only on the boundaries. More specifically, we see two different boundary states; (1) the  surface pair-density-wave state discussed  in \cite{barkman2018surface} (referred to as boundary PDW here), and the boundary superconducting state which can appear in conventional superconductors \cite{samoilenka2019Tc2}. The crosses mark the points for which the pairing field is shown in \figref{fig: 1D BdG}. Remaining parameters are set to $t=1$, $V=2$ and $\mu = 0.5$.} \label{fig: phase diagram BdG}
\end{figure}

At a microscopic level, the   mechanism of formation of boundary states has many complex aspect beyond the  simple energy argument given in \cite{barkman2018surface}.  As pointed out in \cite{barkman2018surface}, at the level of Ginzburg-Landau theory, the state has oscillatory energy density and boundaries are accompanied with beneficial energy segments. However the ability of the system to start with such a segment depends on microscopic boundary conditions applied at a single-electron level.
The energetic argument  clearly indicates that Caroli-de Gennes-Matricon (CdGM) boundary conditions 
that conjecture zero normal derivative of the gap modulus \cite{deGennes_Boundary,CdGM_french,CdGM_Coherence,deGennes_superconductivity} should 
not be expected to hold for pair-density-wave case because the system has clear energetic
preference for having high gradients of the gap modulus near the boundary.
In fact that the Caroli-de Gennes-Matricon (CdGM)  boundary conditions  do not hold   even 
for ordinary superconductors, at least for clean surfaces, as was
recently  shown microscopically in \cite{samoilenka2019Tc2}. It leads to the fact that for the simplest non-FFLO superconductors, there is an enhancement of the superconducting gap at the boundaries, and superconductivity of clean surfaces  \cite{samoilenka2019Tc2}.
Therefore by varying the fermionic imbalance in the BdG formalism,
one can investigate how the surface pair-density-wave states are
connected with the boundary states in non-FFLO regime discussed in  \cite{samoilenka2019Tc2}.
By varying both the temperature and the spin-population imbalance, we obtained the complete phase diagram shown in \figref{fig: phase diagram BdG}.
The phase diagram shows both the PDW boundary state regime and the non-PDW boundary state regime.

\begin{figure}[t]
	\center
	\includegraphics[width=0.99\linewidth]{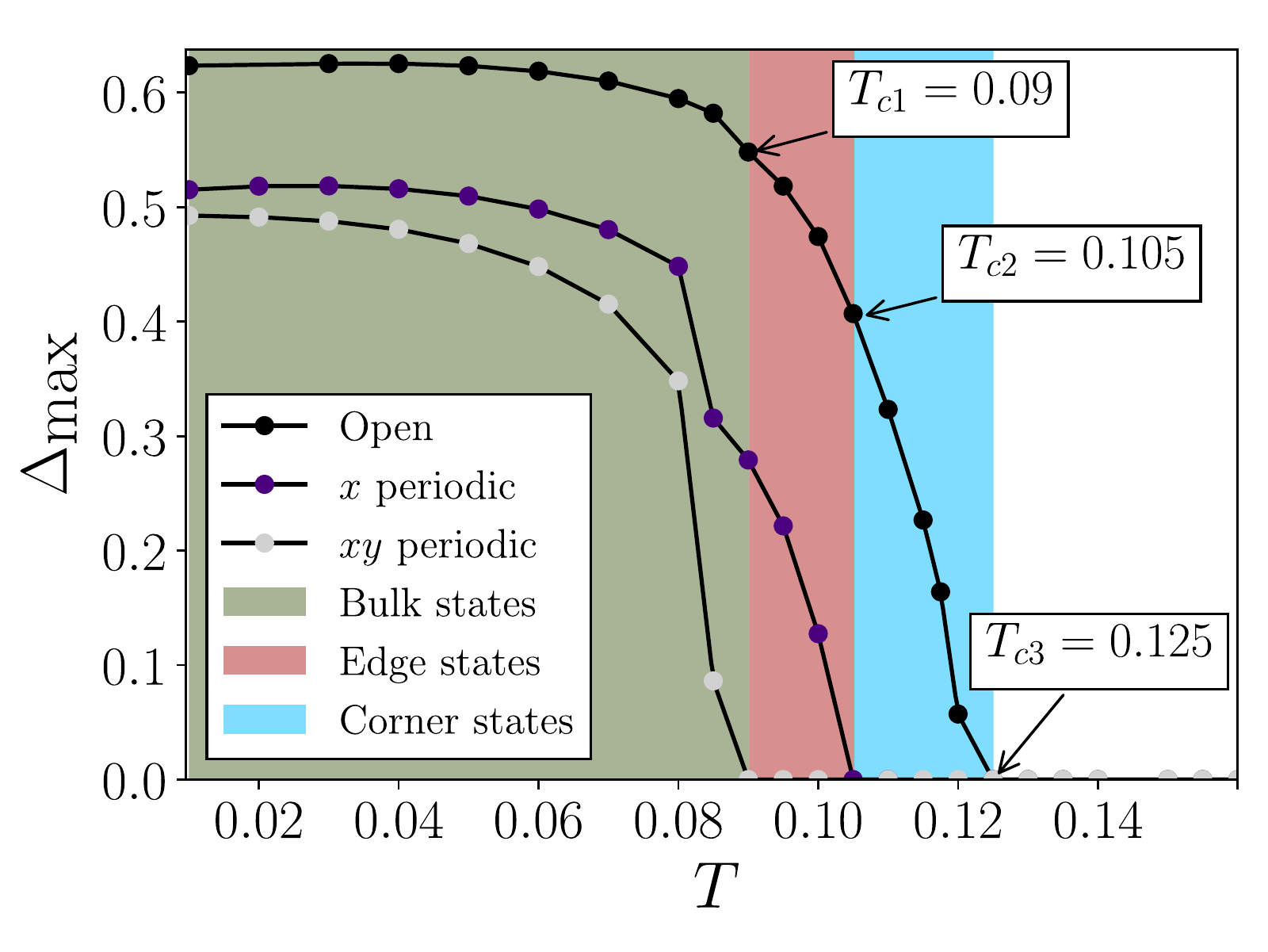}
	\caption{
		Maximum absolute value of the pairing field $\Delta$ obtained numerically in BdG model \eqref{eq: Mean-field Hamiltonian} in two dimensions, for various temperatures $T$ at fixed spin-population imbalance $h=0.4$. Remaining parameters are set to $V=2.5$, $\mu = -0.4$ and $t=1$. The simulations are performed for system with: periodic boundary conditions in both directions $x$ and $y$ (gray-dotted line), which gives $T_{c1} = 0.09$, periodic boundary conditions in only $x$ direction (purple-dotted line) yielding $T_{c2} = 0.105$ and finally open boundary conditions (black-dotted line), leading to $T_{c3} = 0.125$.
	}\label{fig:2D-BdgMaxDelta}
\end{figure}

\begin{figure}[t]
	\center
	\includegraphics[width=0.99\linewidth]{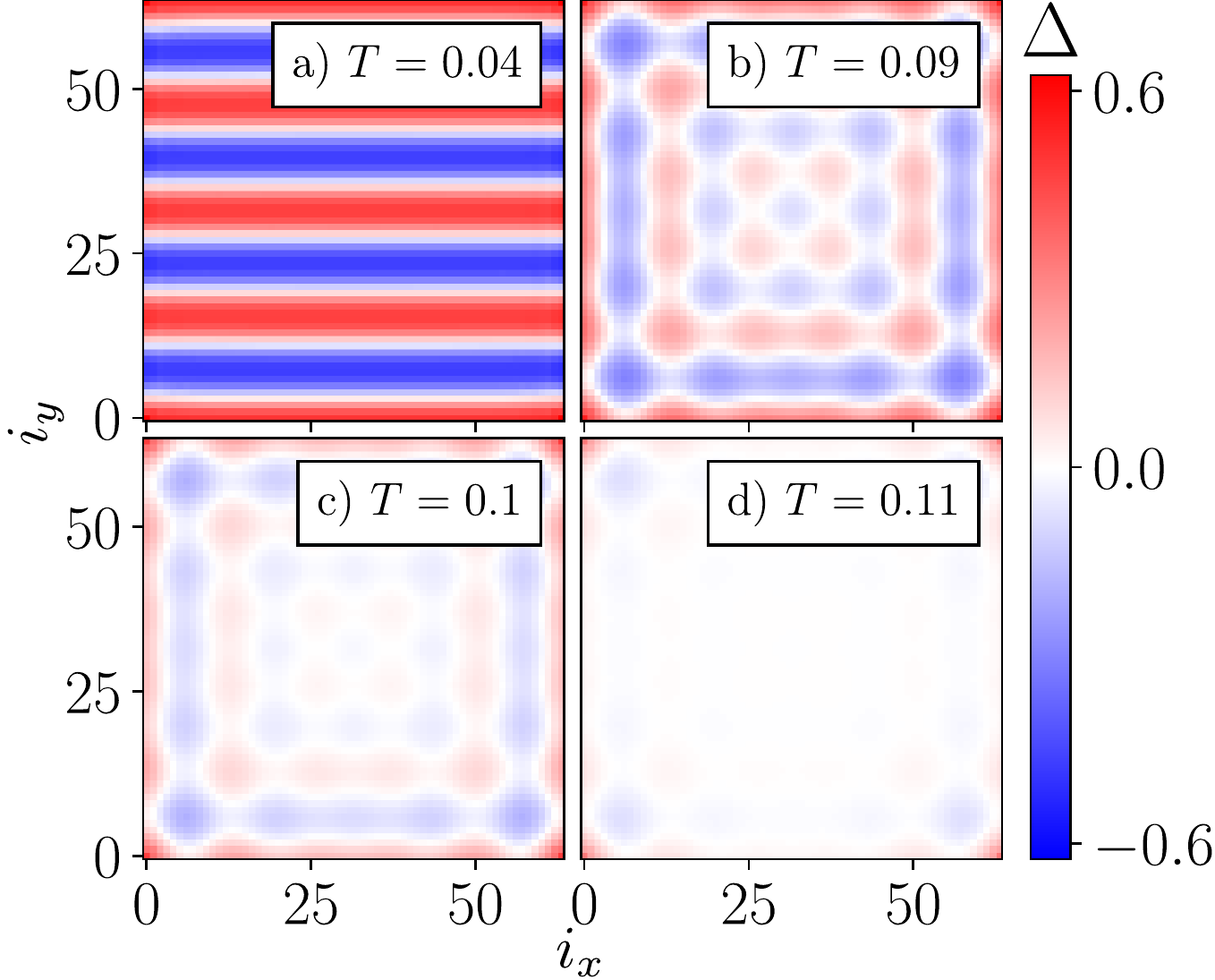}
	\caption{
		Pairing field $\Delta$ obtained numerically in BdG model \re{eq: Mean-field Hamiltonian} with open boundary conditions. The panels display: \textbf{a)} typical LO bulk modulation \textbf{b,c)} the edge pair-density-wave state: the pairing field is enhanced along the boundary while it decays to zero in the bulk, by an exponentially dampened oscillation. \textbf{d)} corner state with pairing field enhanced in corners, which decays also along the boundary. The remaining parameters are fixed to $h=0.4$, $V=2.5$, $\mu = -0.4$ and $t=1$.
	}\label{fig:2D-Bdg_panels}
\end{figure}

\begin{figure}[t]
	\center
	\includegraphics[width=0.99\linewidth]{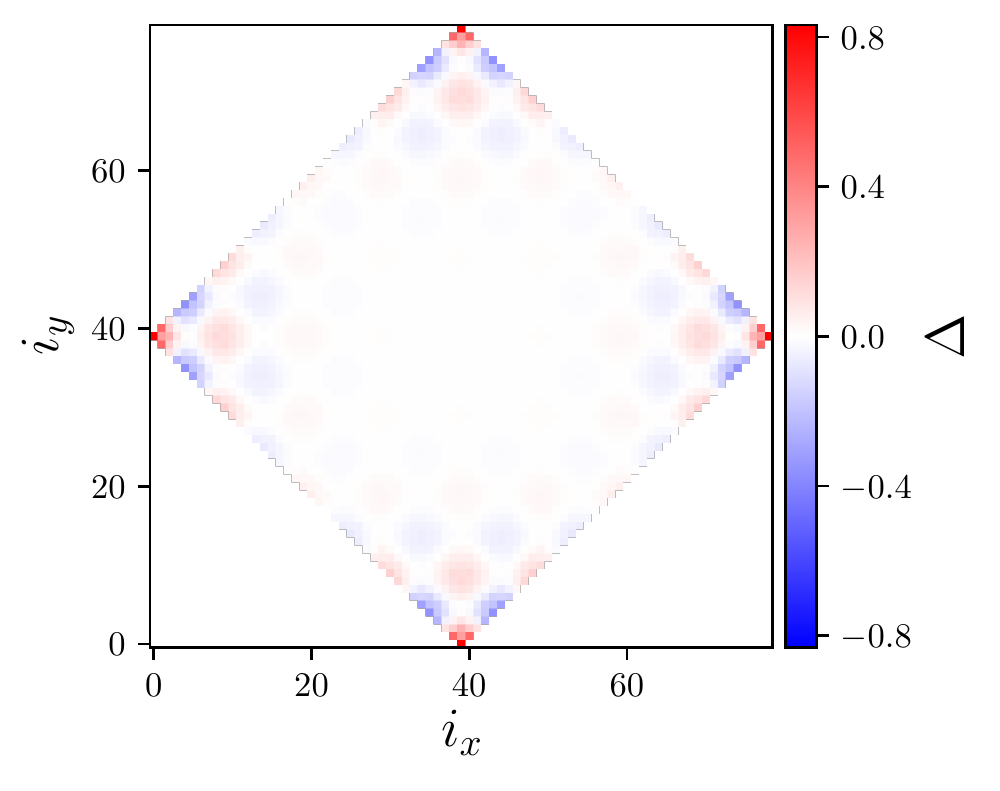}
	\caption{
		Pairing field $\Delta$ obtained numerically in BdG model \re{eq: Mean-field Hamiltonian}. The state occurs for temperature  $T=0.05$, and spin-population imbalance $h=0.5$. The remaining parameters are set to $V=2.5$, $\mu = -0.4$ and $t=1$.
	}\label{fig:2D-BdgSS_along}
\end{figure}

Next we show the existence of the boundary states in two dimensions by solving the BdG equation in \eqref{eq: selfconsistent}.
Obtained pairing fields are shown in \figref{fig:2D-BdgMaxDelta}, \figref{fig:2D-Bdg_panels} and \figref{fig:2D-BdgSS_along}.
Consideration of two-dimensional samples are computationally much more demanding than one-dimensional.
Hence we do not perform full phase diagram exploration, but demonstrate boundary states by fixing the spin-population imbalance $h = 0.4$ and gradually increase the temperature, see \figref{fig:2D-BdgMaxDelta}.
First, we study a square system which is periodic in all directions -- in this case no boundary states can form and we obtain bulk critical temperature $T_{c1}$ by identifying the temperature at which the maximal pairing field goes to zero.
Next, for a sample with periodic boundary conditions only in one direction, edge states are formed, which gives edge critical temperature $T_{c2}$, see \figref{fig:2D-BdgMaxDelta}.
Finally, for open boundaries corner states survive even for higher temperatures, until the corner critical temperature $T_{c3}$ is reached.

For selected values of $T$ we plot configurations of order parameter in \figref{fig:2D-Bdg_panels}. The state in \figref{fig:2D-Bdg_panels}a is the usual LO modulation in one direction. States in \figref{fig:2D-Bdg_panels}b, c are identified as the edge pair-density-wave, in which the pairing field is enhanced along the sample edges and decays as an exponentially dampened oscillation into the bulk. As temperature is increased, the pairing field vanishes on the edges but remain non-zero in the corners, as shown in \figref{fig:2D-Bdg_panels}d. Finally, at even higher temperatures, the gap field vanishes completely and the system transitions to the normal state. These transitions would be the two-dimensional analogue of the multiple phase transitions that occur in the three-dimensional cube in  \figref{fig: 3D Ginzburg-Landau simulations}.
The BdG simulations confirm the existence of the boundary states.

Note however, that the precise structures of the edge and corner states obtained from the BdG calculations differ from those predicted by GL theory, which is not surprising given that they correspond to different physical systems.
Indeed, BdG is considered in tight binding limit and hence corresponds to localized electrons with much stronger interactions than free electron approximation conventionally used to obtain GL model.
The microscopic BdG results in \figref{fig:2D-Bdg_panels} exhibit parallel modulation and exponential decay, while the Ginzburg-Landau results in \figref{fig: 3D Ginzburg-Landau simulations} indicate modulation along the edges. The apparent conclusion obtained from both simulations is the existence of the edge and corner states and sequence of phase transitions.

On the other hand, when replacing the Laplacian term by $\sum_{ij}|\partial_i \partial_j \psi| ^2$ we observe a strong resemblance between the GL and BdG simulations - compare for example we obtain configuration of the order parameter in GL, which is very close to that of BdG: compare \figref{fig: new Laplacian regularization}b and \figref{fig:2D-Bdg_panels}c.
In this case modulation of $\psi$ is mostly orthogonal to the boundary.
However, the modification of GL model with the Laplacian term  replaced by $\sum_{ij}|\partial_i \partial_j \psi| ^2$, in contrast to the BdG solutions, does not have the phases with gap formation on the edges and vertices in the absence of the gaps on the faces.

To demonstrate the presence of edge-modulated states on the BdG level we consider a sample with edges forming an angle of $\pi/4$ with the crystalline axis. The effect of this choice influences the interactions of the edge sites.
In this case there appears modulation of the gap field along the boundaries. Figure \ref{fig:2D-BdgSS_along} shows this result, characterized by the order parameter suppression in the bulk and enhancement in the corners.

In context of cold atoms a phase separation between FFLO-like bulk and normal or BCS-like boundary was predicted in a one-dimensional harmonic trap
 \cite{FFLO_N_BCS_Orso,FFLO_N_BCS_Hu_Liu_Drummond,FFLO_N_BCS_Hu_Liu_Drummond_non_0_T}.
Note, that these states appear as different phase due to spatially variating local density approximation chemical potential $\mu(z) = \mu_0 - V(z)$ inside the trap
making the system form different states at different distances from the trap center.
Where the potential is deep, i.e. in the center of a trap, FFLO state appears while
near the boundary of the trap, normal or BCS states are formed.
By contrast, the states that we find not only 
are represented by a very different solution but appear under different conditions: induced by hard wall potential, when particles are not permitted to hop outside the sample, which is equivalent to infinite chemical potential at a boundary.
It does not have 
a direct counterpart in the phase separation picture
in traps.
Another principal difference is that our states 
rely on the density of state structure
near   boundaries, as can be most clearly seen in a BCS superconductor with zero population imbalance \cite{samoilenka2019Tc2}.

\section{Conclusions}
In conclusion, we considered boundary effects in a superconductor where Cooper pairing
involves electrons with finite center-of-mass momentum (Fulde-Ferrel-Larkin-Ovchinnikov instability).
Firstly, we demonstrated that the standard microscopically-derived Ginzburg-Landau model cannot be
used to describe such superconductors in the presence of boundaries in dimension larger than one since it gives a spurious divergence 
of the free energy near the boundaries. To describe such states we constructed a generalized functional with higher order derivative terms were retained to guarantee finiteness of the free energy density.
We showed that at the mean-field-level the system undergoes a sequence of the phase transitions where
each transition is associated with  decreased dimensionality of the superconducting state.
Namely as temperature is increased the system first looses bulk superconductivity but retains two-dimensional superconductivity on its faces.  The next phase transition is associated with the loss of two-dimensional superconductivity on the faces but retaining one-dimensional superconductivity on the edges. When the temperature is increased further the superconducting gaps survives only at the vertices.
The transition from the bulk to surface pair-density-wave superconductivity is also 
captured by an alternative and simpler form of the Ginzburg-Landau functional described in \secref{sec: no divergence GL}.

In order to demonstrate  the existence of multiple phase transitions in a model that does not rely 
on Ginzburg-Landau expansion we solved numerically Bogoliubov-de-Gennes model in two dimensions.
Clearly solution demonstrate the phase transition from the bulk to boundary superconductivity.

The results provide a new route to identify and study properties of pair-density-wave superconductors.
The results predict a discrepancy between the bulk-dominated specific heat identification of critical temperature and
surface-dominated current and diamagnetic response probes. The predicted possibility of a  different modulation
of the order parameter on the surface also has direct implication for scanning tunneling microscopy and Josephson experiments. 

The current work also opens up the question of the physics of boundary states beyond mean-field
approximation. Beyond mean-field approximation, quasi-one-dimensional superconductors
are always phase disordered in thermodynamic limit at any finite temperature. However, again as 
in any superconducting wire that statement applies only to thermodynamic limit and it's expected
to see the mean-field sequence of the phase transitions in local probes such as scanning tunneling microscopy.
The surface phase transition beyond mean-field approximation should be of the Berezinskii-Kosterlitz-Thouless type.
However multiple transitions are also possible beyond mean-field in cases where the modulation is along the surface, associated
with melting of stripes and loss of crystalline modulation order. This will be a
counterpart of  multiple transitions beyond mean-field
approximation that attract currently substantial interest in the conventional two-dimensional Larkin-Ovchinnikov cases \cite{Agterberg2008,Berg2009} that yields additional paired and charge-$4e$ phases.
Our results indicate that beyond mean-field approximation, the charge  charge-$4e$ superconductivity via  Berezinskii-Kosterlitz-Thouless type transition can occur also at the surfaces of bulk superconductors (for a detailed general discussion of the fluctuation-induced paired phases in low-dimensional multicomponent superconductors see \cite{babaev201547phase,svistunov2015superfluid}).

\begin{acknowledgments}
We thank Robert Vedin for his valuable assistance.
The work was supported by the Swedish Research
Council Grants No. 642-2013-7837, 2016-06122, 2018-03659 and G\"{o}ran Gustafsson  Foundation  for  Research  in  Natural  Sciences  and  Medicine. A part of this work was performed at the Aspen Center for Physics, which is supported by National Science Foundation grant PHY-1607611
We gratefully acknowledge the support of NVIDIA Corporation with the donation of the Quadro P6000 GPU used for this research. Published with the support from the L\"{a}ngman Culture Foundation.
\end{acknowledgments}

\section*{Appendix}
It is common to use variational approach to derive BdG equations \cite{deGennesBook}.
Namely, the self consistency equations are obtained by setting the variation of the free energy $\Omega$ with respect to mean fields to zero.
\begin{equation}
\Omega = \tr\left[ \mathcal{H} \rho_{MF} + T \rho_{MF} \ln \rho_{MF} \right]
\end{equation}

where $\mathcal{H}$ is Hamiltonian corresponding to exact Fermi-Hubbard model \eqref{FermiHubbard} and $\rho_{MF} = \frac{e^{-H_{MF}/T}}{\tr e^{-H_{MF}/T}}$ with mean field Hamiltonian $H_{MF}$.

The main difference from our derivation in \secref{sec: BdG approach} is that one has to include additional mean fields in $H_{MF}$ via Hartree term  $\sum_{i, \sigma} U_{i \sigma} \psi^\dagger_{i \sigma} \psi_{i \sigma}$.
It amounts to replacing matrix $H$ in self consistency \eqref{Delta_self_consistent} for $\Delta_i$ with $M = H + U$, where
\begin{equation}
U_{i j} = 
\begin{pmatrix}
\delta_{i j} U_{i \uparrow} & 0 \\
0 & - \delta_{i j} U_{i \downarrow}
\end{pmatrix}
\end{equation}

In addition, the fields $U_{i \sigma}$ have to be solved for self consistently using
\begin{eqnarray}\label{Hartree_self_consis}
\begin{aligned}
U_{i \uparrow} & = - V \vec{h}_i^T f(- M) \vec{h}_i = -V\langle n_{i\downarrow}\rangle, \\
U_{i \downarrow}&  = - V \vec{e}_i^T f(M) \vec{e}_i = -V\langle n_{i\uparrow}\rangle.
\end{aligned}
\end{eqnarray}

\begin{figure}[H]
\center
\includegraphics[width=0.49\textwidth]{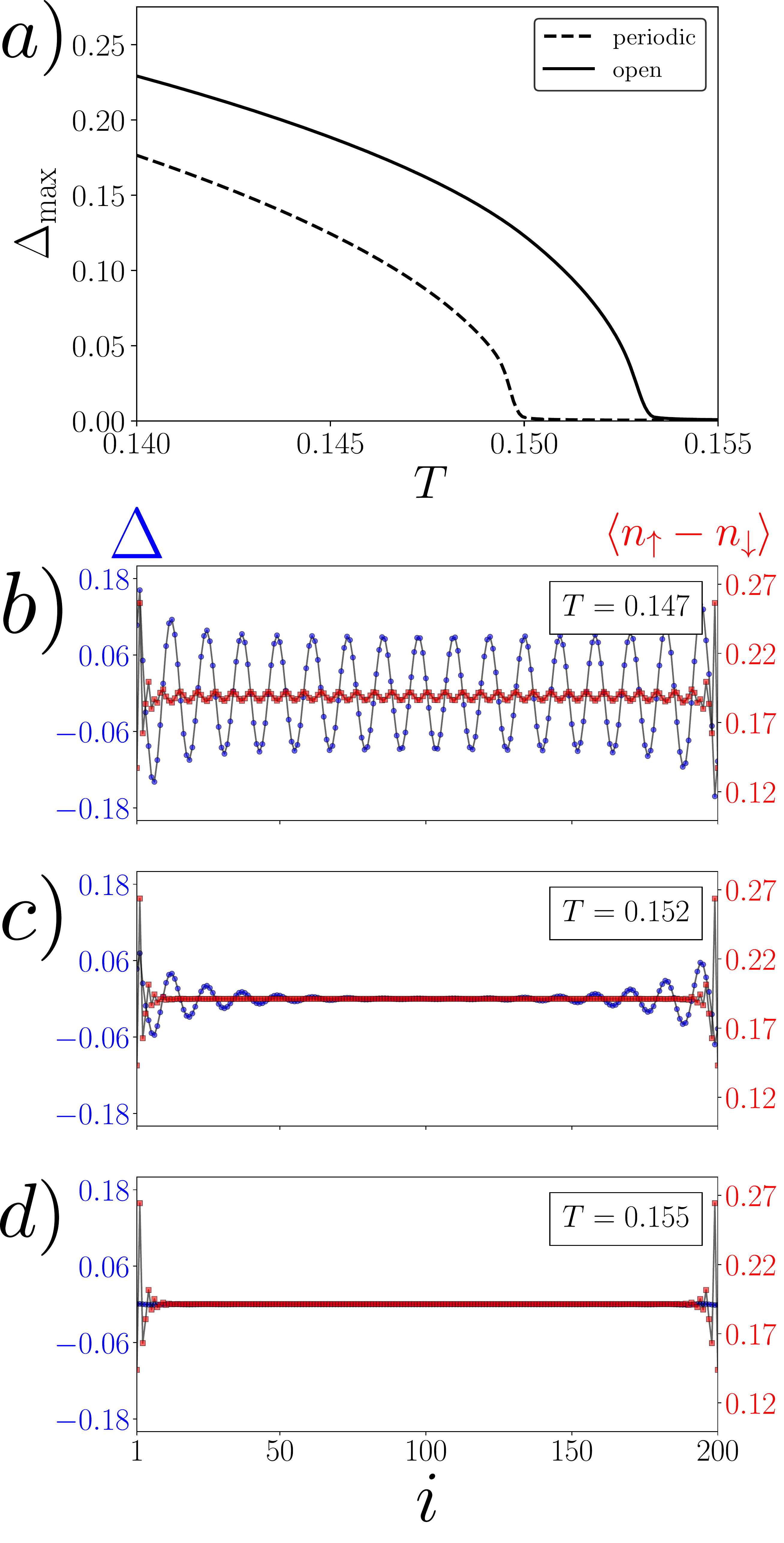}
\caption{
	\textbf{a)} Comparison between the maximal value of pairing potential $\Delta$ for various temperatures, using either open (full line) or periodic (dashed line) boundary conditions in one dimension.
	Boundary states are present at $T \in [0.150, 0.153]$.
	Pairing potential $\Delta$ (blue) and spin population imbalance $\langle n_\uparrow - n_\downarrow \rangle$ (red) of:
	\textbf{b)} LO state at $T=0.147$, which exhibits modulation both in $\Delta$ and $\langle n_\uparrow - n_\downarrow \rangle$, with the later one being much less prominent.
	\textbf{c)} modulated boundary state with exponentially dampened oscillations at $T=0.152$.
	\textbf{d)} state at which superconductivity is lost while the boundary modulation in the spin imbalance remains at $T=0.155$.
	Remaining parameters are set to $t=1$, $\mu=-0.65$, $h=0.7$ and $V=2.5$.
}\label{fig:Hartree1d}
\end{figure}

In \figref{fig:Hartree1d} we present results in one dimension which incorporate self-consistent solution for both the pairing field and the Hartree fields, for various temperatures, keeping the remaining parameters fixed. We show result using both periodic and open boundary conditions, in order to investigate the existence of boundary states.
We find that boundary states also exist in this model but with a smaller difference between the  critical temperatures.
Note, that modulation in spin population imbalance $\langle n_\uparrow - n_\downarrow \rangle$ is much smaller and is twice as frequent as that of $\Delta$.
The later is easy to explain by the fact that each of the populations $\langle n_\sigma \rangle$ in \eqref{Hartree_self_consis} is even function of $\Delta$.
Hence they depend on $|\Delta|$, which has twice shorter period than $\Delta$. Note that the spin-population imbalance modulation is enhanced close to the boundary, but the decay length significantly shorter than for the pairing potential. This enhancement remains even for higher temperatures, when superconductivity is completely lost and is related to Friedel oscillations.

\bibliographystyle{apsrev4-1}

%

\end{document}